\documentclass[aip,jap,reprint,10pt,showpacs,unsortaddress,longbibliography]{revtex4-1}
\usepackage{graphicx}
\usepackage{amsfonts}
\usepackage{amsmath}
\usepackage{natbib}
\usepackage{xcolor}

\begin{document}

\title{Exploring a new ammeter traceability route for ionisation chamber measurements}

\author{S.~P.~Giblin}
\email{stephen.giblin@npl.co.uk}
\affiliation{National Physical Laboratory, Hampton Road, Teddington, Middlesex TW11 0LW, United Kingdom}
\author{G.~Lorusso}
\affiliation{National Physical Laboratory, Hampton Road, Teddington, Middlesex TW11 0LW, United Kingdom}
\affiliation{Department of Physics, University of Surrey, Guildford, GU2 7XH, United Kingdom}

\date{\today}

\begin{abstract}
We compared the performance of a commercial ammeter and a home-made integrating electrometer in reading ionisation chamber currents less than 100~pA. The integrating electrometer charges a capacitor with the unknown current and measures the resulting rate of change of voltage, whereas the ammeter uses a high-value resistor as the feedback element to an amplifier which converts current to voltage. The noise performance of both systems was very similar for averaging times less than 1000 seconds. Both systems were calibrated using a reference current source with 1 part per million (ppm) accuracy, revealing an error of 460 ppm in the electrometer indicated current, of unknown origin. This error is well within the uncertainty budget for radionuclide calibrations, but much larger than the individual uncertainties in the traceable calibrations of capacitance, voltage and time. The noise in the ionisation chamber current was much larger than the noise floor of both instruments, with tests providing strong indication that the excess noise originated in the high voltage source used for energising the chamber.

\end{abstract}

\pacs{1234}

\maketitle

\section{\label{IntroSec} Introduction}

Ionisation chambers are of great utility for measuring radionuclide activities and half-lives. The chamber outputs a current proportional to the activity of the source inside the chamber, with the constant of proportionality determined by primary calibration methods involving absolute counting of decay events from a diluted source \cite{schrader1997activity,schrader2007ionization}. The linearity and stability of the ionisation chamber current measurement is ensured by traceable calibration of the current measuring instrument. Historically, these instruments have usually been capacitor-ramp electrometers which integrate the ionisation chamber current and allow the current to be calculated according to $I = C\frac{dV}{dt}$. For ionisation chamber currents in the picoamp to nanoamp range, voltage ramp rates of $\frac{dV}{dt} \sim 1$~V/s require capacitances $C$ in the picofarad to nanofarad range. Such capacitors are available commercially as low-loss air or sealed-gas units possessing long-term stability at the part-per-million level, and low sensitivity to temperature and humidity changes. The relevant calibrations of voltage, capacitance and time are available as standard services from national metrology institutes (NMIs), and accredited laboratories, with relative uncertainties less than $10$ parts per million (ppm), and in the absence of complicating factors these low uncertainties are transferred directly to the measured current.

In the last 15-20 years, a number of developments have occurred in the field of small current metrology which encourage a fresh look at ionisation chamber current readout methods. In response to industry demand, a number of NMIs have inaugurated calibration services for nanoamp-level ammeters with uncertainties as low as $\sim 10$~ppm. Reference currents are usually sourced by applying a linear voltage ramp to a low-loss capacitor (essentially the reverse process of a capacitor-ramp electrometer) \cite{willenberg2003traceable,van2005accurate,fletcher2007new,callegaro2007current}. To validate these new services, the first international intercomparison of reference current sources was undertaken \cite{willenberg2013euromet}. While broadly validating NMI capability, the comparison could not provide information at uncertainty levels much below $\sim 100$~ppm due to transport instability and environmental effects in the commercial ammeters used as transfer standards. In parallel, research into prototype quantum current sources, known as electron pumps, which generate small currents by moving electrons one at a time \cite{pekola2013single}, focused attention on small-current metrology at the lowest possible uncertainty level. In this research setting, currents of order $100$~pA have been measured with combined uncertainties of $\sim 0.2$~ppm \cite{stein2016robustness,zhao2017thermal}. A practical spin-off from the electron pump research has been the ultrastable low-noise current amplifier, or ULCA \cite{drung2015ultrastable,drung2017ultrastable}. This instrument, following calibration using a cryogenic current comparator (CCC)\cite{drung2015improving}, can either source or measure small currents with uncertainties as low as $0.1$~ppm, and has demonstrated stability under international transportation at the $1$~ppm level \cite{drung2015validation}. Recently, different versions of the ULCA have been tested, including ones with high gain and small, stable, offset suitable for the measurement of the very low background currents from ionisation chambers \cite{drung2017ultrastable,krause2017measurement}.

Inspired by these developments, in this paper we test an alternative traceability route for ionisation chamber currents: an ammeter calibrated directly using a primary reference small-current source. We compare this ammeter method with an established capacitor-ramp method in which the traceability is to standards of capacitance, voltage and time, and discuss the advantages and limitations of each. We also address an important and neglected question in ionisation chamber metrology: how does the random uncertainty in the measured current depend on the measurement time, and what is the optimum interval between chamber background measurements.

\section{\label{TracSec} Traceability routes}

\begin{figure}
\includegraphics[width=8.5cm]{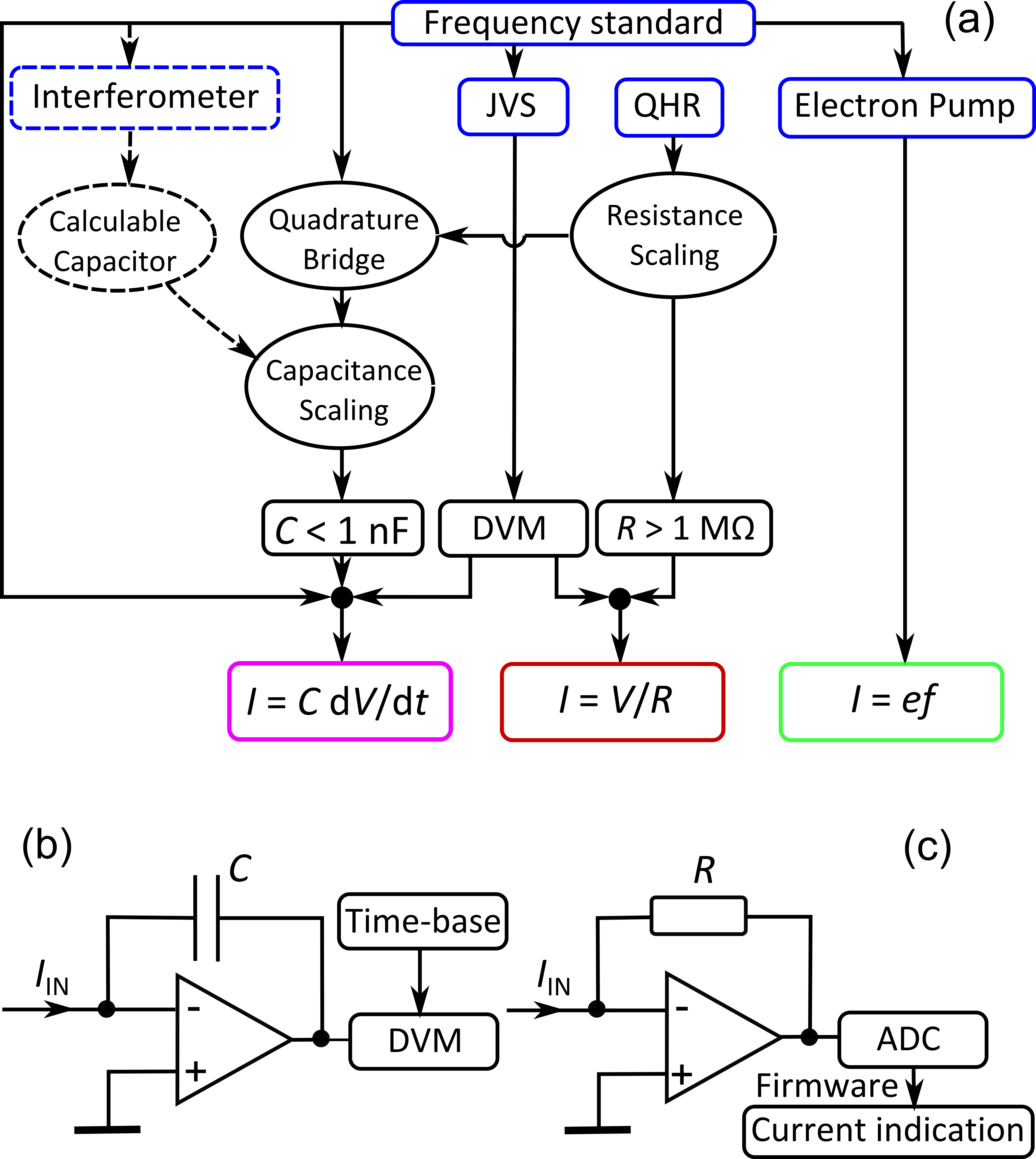}
\caption{\label{TraceFig}\textsf{(a): Diagram showing routes for traceable generation of small currents via three main mechanisms: capacitor ramping, Ohms law and the controlled transport of charge. Abbreviations are JVS = Josephson voltage standard, QHR = quantum Hall resistance, DVM = digital voltmeter. The elementary charge is denoted $e$. (b): Simplified schematic circuit diagram of an integrating electrometer. (c): Simplified schematic circuit diagram of a feedback ammeter.}}
\end{figure}

In figure \ref{TraceFig} (a), three complete traceability routes for small electrical currents are summarized, starting with primary standards at the top. The electron pump is included for completeness; although they currently have the status of research devices, electron pumps offer a very direct tracebility route and are likely to play a role in primary current metrology in the future \cite{pekola2013single}. In this paper, we will be concerned mainly with the first two - the capacitor ramp method and the resistor/voltage method.

The capacitor ramp method realizes current via the rate of change of voltage across a capacitor, and the concept can be applied to either the generation or measurement of a current. The traceability route for capacitance is either to the dc quantum Hall resistance (QHR) via a quadrature bridge and ac/dc transfer resistor, or via the calculable capacitor, which realizes a small ($\lesssim 1$~pF) capacitance based on a length measurement. Both these routes are moderately complex to implement, but the end result is that standard capacitors of $1$~nF or less can be calibrated routinely at audio frequencies with uncertainties of order $1$~ppm. Voltage is traceable to the ac Josephson effect, and digital voltmeters (DVM's) can be calibrated directly against a Josephson voltage standard (JVS), or indirectly using a calibrator or a Zener diode voltage reference. High-specification DVMs may drift by at most a few ppm in a 1-year calibration interval and have non-linearity errors less than $1$~ppm. The third traceable quantity, time, can be realized with ppm accuracy in a number of ways - for example using a commercial off-air frequency standard. The quantity $C\times \frac{dV}{dt}$ can consequently be realized with an uncertainty of a few parts per million, and precision reference current sources have mostly used this route\cite{willenberg2003traceable,van2005accurate,fletcher2007new,callegaro2007current}. 

Generation of sub-nA reference currents using a resistor and voltage source is less common. This may be because high-value standard resistors, in contrast to sub-nF air-gap capacitors, can have temperature co-efficients as large as few tens of ppm per degree, and therefore require additional environmental control to reach ppm-level accuracy. Calibration uncertainties of high-value resistors have also been generally higher than low-value capacitors, although ppm-level calibration uncertainties of resistors up to $1$~G$\Omega$ are now attainable using CCCs \cite{fletcher2000cryogenic,bierzychudek2009uncertainty}. The ULCA \cite{drung2015ultrastable} also generates and measures current with respect to an internal $1$~M$\Omega$ resistor and an external DVM, and as already noted, has demonstrated 1-year stability at the ppm level. The resistor and voltage source method has the obvious advantage that current can be generated continuously without being constrained by a capacitor charge-discharge cycle.

A problem with the capacitor ramp method is that the low calibration uncertainties of the standard capacitors are achieved using voltage-transformer bridge techniques\cite{kibble1984coaxial} which work at audio frequencies. Calibrations are typically performed at $1$~kHz, and the techniques can be extended down in frequency to a practical lower limit of $\sim 25$~Hz. In contrast, capacitor ramp methods for generating or measuring small currents operate at frequencies many orders of magnitude lower, in the millihertz range. One study found that some samples of standard capacitor exhibited unexpectedly large frequency dependence in the range $\sim 10$~mHz - $1$~kHz, up to several hundred ppm \cite{giblin2010frequency}, which is certainly far in excess of the $1$~kHz calibration uncertainty and begins to impact the uncertainty budgets of NMI-level ionisation chamber readout systems. Either the capacitance needs to be measured at the ramp frequency, which is a laborious and non-standard procedure\cite{giblin2010frequency}, or the capacitance uncertainty must be expanded to allow a worst-case scenario. This issue reduces somewhat the apparent advantages of the capacitor ramp method, and prompts fresh consideration of the resistor and voltage method.

\section{\label{SystSec}Current measurement and generation systems}

\subsection{Current measurement systems}

The two types of current readout system investigated in this paper, the capacitor ramp electrometer and the feedback ammeter, are illustrated schematically in figure \ref{TraceFig} (b,c). We will refer to them subsequently as the `electrometer' and the `ammeter' respectively. Both types of instrument use a high-gain amplifier with feedback; the feedback element is a capacitor in the case of the electrometer, and a resistor in the case of the ammeter. The electrometer used in this study employs a home-made amplifier with an external integrating air-gap capacitor of value $\sim 500$~pF, and an external DVM (Datron model 1061) triggered with a calibrated $1$~s interval between readings. We denote the current measured by the electrometer $I_{\text{E}} \equiv C_{\text{corr}} \times \frac{dV}{dt}$. Here, $C_{\text{corr}}=C_{\text{cal}}+C_{\text{stray}}$ where $C_{\text{cal}}$ is the calibrated value of the standard capacitor, and $C_{\text{stray}}$ is the stray capacitance correction. For the bulk of the study, excepting the data of figure \ref{NoiseFig} (f,g), the ammeter was a Keithely model 6430 set to the 1 nA range \footnote{We could have used the 100 pA range, achieving slightly lower instrument noise at the expense of slower response time. However, the noise in $I_{\text{A}}$ is roughly two orders of magnitude higher than the instrument noise floor so no benefit is obtained by using a lower range}. The resistive feedback of the ammeter gives an output voltage $=I_{\text{in}}R$, which is digitized by an analogue-to-digital converter (ADC) internal to the instrument, and converted to a current reading  by the instrument's firmware. The feedback resistor ($\sim 1$~G$\Omega$ on the 1 nA range) is internal to the ammeter, and the ammeter was calibrated by supplying it with a reference current. 

\begin{figure}
\includegraphics[width=8.5cm]{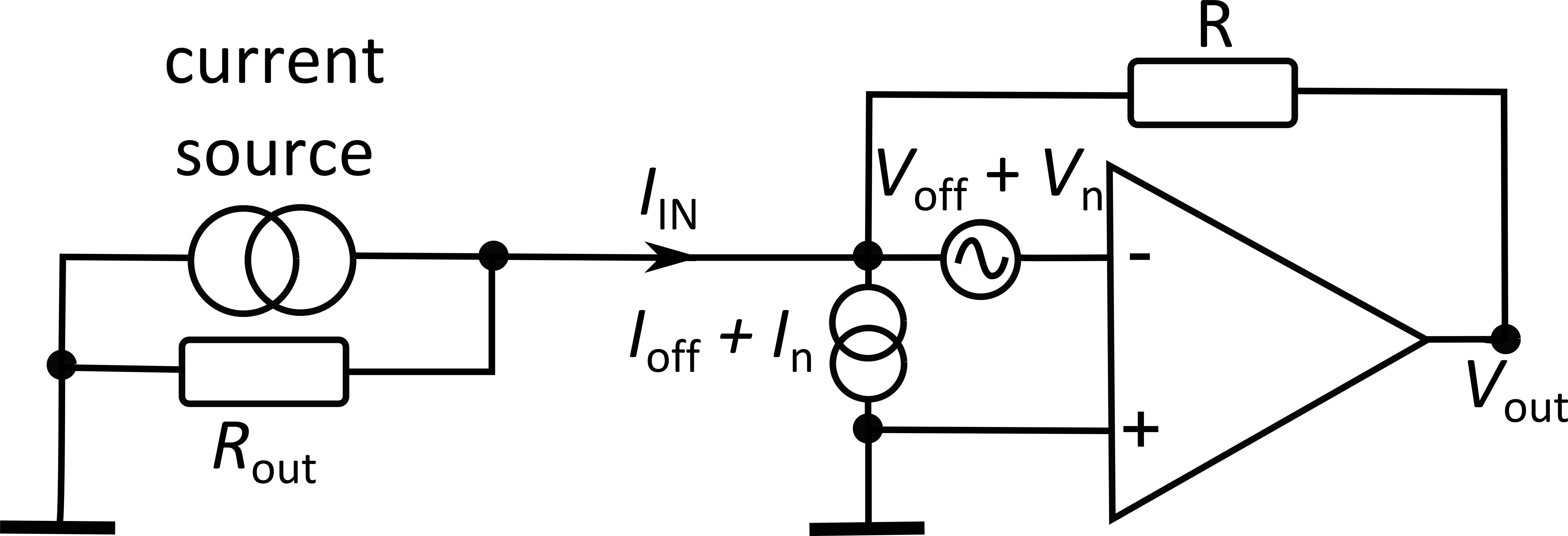}
\caption{\label{CircuitFig}\textsf{Simplified schematic circuit diagram showing the input stage of an ammeter connected to a non-ideal current source with finite output resistance $R_{\text{out}}$. The offset current and voltage are denoted $I_{\text{off}}$ and $V_{\text{off}}$ respectively, and the current and voltage noise are denoted $I_{\text{n}}$ and $V_{\text{n}}$ respectively.}}
\end{figure}

\subsection{noise considerations}

In figure \ref{CircuitFig}, we present an expanded circuit model for the input stage of an ammeter connected to a current source, including the offsets and noise sources\footnote{The term 'noise', frequently used in this paper, means random fluctuations in a measured signal, irrespective of the origin of those fluctuations.} present in real ammeters, and the finite output resistance $R_{\text{out}}$ of the current source. Additional noise due to the current source itself is not considered in this model. The voltage offset and noise are represented by a single source in the diagram for convenience (and likewise for the current offset and noise) but this should not be taken to imply that they are due to the same process or component in the amplifier. The same circuit describes the electrometer, but with $R$ replaced by a capacitor. A detailed discussion of amplifier properties is beyond the scope of this paper, but some qualitative comments will help with interpreting the data of sections \ref{TypeASec} and \ref{AgreeSec}. The total amplifier noise is the sum of three contributions: the current noise $I_{\text{n}}$, the thermal noise in the feedback resistor $R$ (in the case of capacitive feedback, there is no thermal noise), and the voltage noise $V_{\text{n}}$ driving a noise current in the source resistance $R_{\text{out}}$ \cite{graeme1996photodiode}. Crucially, while the first two contributions are independent of $R_{\text{out}}$, the last one increases in inverse proportion to $R_{\text{out}}$. The reference current source used for calibrating the ammeter and electrometer (described in the next paragraph) has $R_{\text{out}} = 1$~G$\Omega$, whereas an ionisation chamber presents a very high output resistance, $R_{\text{out}} \gg 1$~G$\Omega$. We therefore expect $V_{\text{n}}$ to contribute more noise during a calibration of the instrument than when measuring an ionisation chamber current, and depending on the relative size of $I_{\text{n}}$ and $V_{\text{n}}$ (we did not separately measure these for either the electrometer or the ammeter) we may expect to see an increase in the noise when the instrument is connected to the reference current source. Generally, designers of amplifiers have to make trade-offs, and it is difficult to make both $V_{\text{n}}$ and $I_{\text{n}}$ arbitrarily small. We note that the measurement of ionisation chamber currents is an application in which $V_{\text{n}}$ can be relaxed somewhat in a specialised instrument design due to the very high output resistance of the source, to enable the smallest possible $I_{\text{n}}$. A commercial ammeter, on the other hand, may offer smaller $V_{\text{n}}$ and larger $I_{\text{n}}$, in order to yield a reasonable total noise when measuring current sources with a wide range of $R_{\text{out}}$.

The same general comments also apply to the offset current and voltage, $I_{\text{off}}$ and $V_{\text{off}}$; in instrument design there is typically a trade-off between the two. We measured $V_{\text{off}}=5$~mV for the electrometer, and $V_{\text{off}} \sim 0.2$~mV for our Keithley 6430 ammeter unit on the 1 nA range. The large offset exhibited by the electrometer caused a $5$~pA offset current to flow when it was connected to the reference current source for the measurements of section \ref{AgreeSec} A, but the on-off calibration cycle subtracted this offset and measured only the gain factor of the electrometer.

\subsection{reference current source}

Our reference current generator consisted of a calibrated, temperature-controlled $1$~G$\Omega$ standard resistor, an uncalibrated voltage source and a calibrated DVM (model Keysight 3458A). The combined type B uncertainty in the reference current was $\sim 1$~ppm. In discussing calibrations, we need first to distinguish the instrument's gain factor from its offset. We describe the relationship between the true current, $I_{\text{true}}$, and current indicated by the instrument, $I_{\text{Ind}}$, as $I_{\text{true}} = (g \times I_{\text{Ind}}) + I_{\text{offset}}$, where $g$ is the gain factor. Our calibration determines only the gain factor $g$. The offset current $I_{\text{offset}}$ is automatically removed from the background-corrected measurements of activity discussed in section \ref{AgreeSec}, since it is present in the current with and without the radionuclide source in the ionisation chamber. We calibrated the gain factor of the ammeter every 2-3 days during the measurement period, and we denote the current measured by the ammeter, after adjusting the indicated current for the gain factor, as $I_{\text{A}}$. Care was taken not to subject the sensitive ammeter preamp unit to mechanical shock, as previous experience with the EM-S24 small-current inter-comparison \cite{willenberg2013euromet} showed that even small mechanical shocks, such as plugging a cable into the preamp, could change the gain factor by several tens of ppm. Following these precautions, the ammeter calibration factor changed by less than $5$~ppm over $2-3$ weeks. For part of the study, we also used the same reference current source to calibrate the electrometer, as detailed in section \ref{AgreeSec}.

\section{\label{TypeASec} Dependence of type A uncertainty on averaging time}

\begin{figure}
\includegraphics[width=8.5cm]{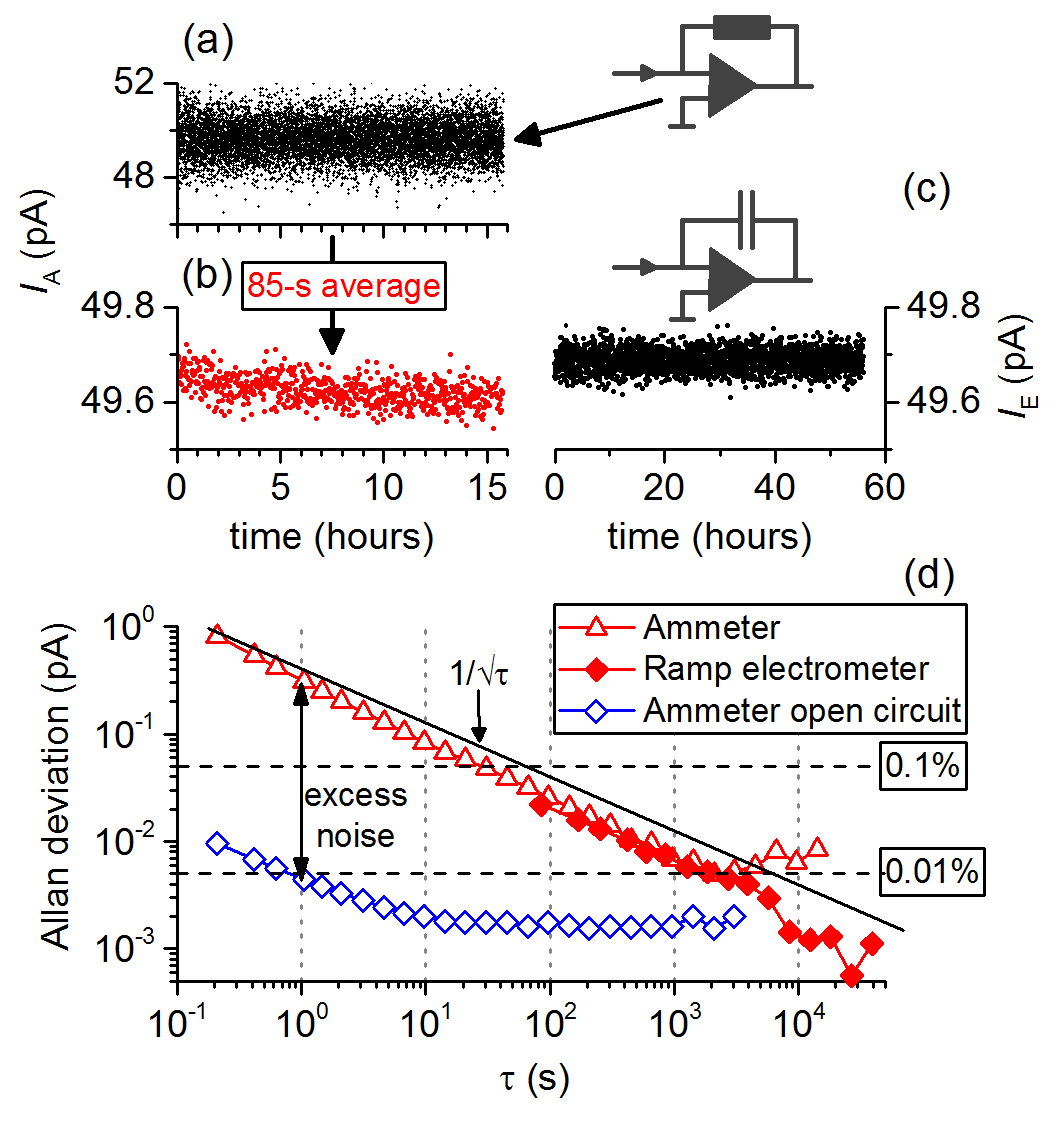}
\caption{\label{AdevFig}\textsf{(a): raw ammeter data obtained while measuring the output of an ionisation chamber containing a Ra-226 source. (b): The data in (a) averaged in 85-second blocks. (c) Data from the same source/chamber combination measured using a capacitor ramp electrometer. Each ramp takes $85$~seconds. (d): Allan deviation as a function of averaging time $\tau$ of the data in plot (a) (open triangles), plot (c) (filled diamonds) and an additional data set obtained with the ammeter disconnected from the ionisation chamber to measure its noise floor (open diamonds). The diagonal solid line is a guide to the eye with slope $1/ \sqrt{\tau}$. The vertical double arrow indicates the difference between the ammeter noise floor and the noise when measuring the ionisation chamber current, and the horizontal dashed lines indicate relative random uncertainties of $0.1 \%$ and $0.01 \%$.}}
\end{figure}

All the radionuclide measurements were performed using the same ionisation chamber, which was of type Vinten 671. To assess the type A (statistical) uncertainty after a given averaging time, we placed a sealed Ra-226 source in the chamber and measured the current for periods of several hours. Raw data from ammeter measurements is shown in figure \ref{AdevFig} (a). The ammeter was set to integrate each data point for $10$ power line cycles (PLC), with the auto zero function disabled, and consequently the raw data set consists of $5$ data points per second. In figure \ref{AdevFig} (b), the same data has been block-averaged so that each plotted point is averaged over $85$ seconds of measurement time. A plot of the ionisation chamber current from the same Ra-226 source, measured using the electrometer, is shown in figure \ref{AdevFig} c. In this plot, each data point is obtained from one voltage ramp cycle. The ramp cycle lasted $85$~s, so the data points in figures \ref{AdevFig} b,c can be directly compared, i.e. each data point corresponds to the instrument integrating the current signal for the same amount of time. The offset of $\sim 0.1$~pA between the two instruments is not significant because these measurements are not corrected for the background, and we will investigate the agreement between the two systems in section \ref{AgreeSec}. The significant feature visible in these long data sets is that the average current measured by the ammeter appears to drift with time, decreasing by $\sim 50$~fA over the first few hours and continuing a downward drift more slowly for the remainder of the measurement time. The rapid drift visible at the start of this data set was rather atypical of the performance of this instrument, and was not the result of mechanical shock. The instrument was powered up and running in acquisition mode for several days prior to the start of the data set. We cannot rule out the possibility that the drift is due to a change in the ambient temperature, coupled to a temperature dependence of the ammeter's gain-setting resistor, but this is unlikely as calibrations of the ammeter spread over two weeks showed the gain to the stable at the $10^{-5}$ level. In contrast to the ammeter data, the current measured by the electrometer appears to be stationary in time. Next, we employ the Allan deviation to more quantitatively investigate this observation.

The Allan deviation is a statistical tool developed as a way of assigning a meaningful statistical uncertainty to data with a non-stationary mean \cite{allan1987should}. It is widely used in time and frequency metrology, and its use in electrical metrology is becoming more widespread, for example to characterize the stability of voltage standards \cite{witt2000using} and current comparator bridges \cite{williams2010automated,drung2015improving}. Here, we briefly summarize it. The Allan deviation $\sigma _{\text{A}}$ is computed from a time-series of data points evenly spaced over a total time $T$. The computation yields $\sigma _{\text{A}}$ as a function of averaging time $\tau$, for $\tau \lesssim T/4$. For the case of frequency-independent noise, $\sigma _{\text{A}}(\tau) = \sigma / \sqrt{\tau}$, where $\sigma$ is the standard deviation of the data; in other words, the Allan deviation is equal to the standard error of the mean, and decreases as the square root of the measurement time. However, in the presence of frequency-dependent noise, the standard error of the mean is no longer a meaningful measure of the type A uncertainty. Two examples of frequency-dependent noise are $1/f$ noise, in which the Allan deviation is independent of $\tau$, and random-walk, or $1/f^2$ noise, in which the Allan deviation increases as the square root of $\tau$. 

The Allan deviation of the time-domain data from figure \ref{AdevFig} (a) and (c) is shown in figure \ref{AdevFig} (d). Note that the first data point for the electrometer is at $\tau=85$~s, the time for one integration ramp, whereas the ammeter data starts at $\tau=0.2$~s, the time to acquire one reading. It is clear that both instruments have very similar $\sigma _{\text{A}}$ for $\tau < 2000$~s, and that $\sigma_{\text{A}} (\tau) \propto 1/ \sqrt{\tau}$. For $\tau > 2000$~s, the behavior of the two instruments diverges. The ammeter enters a regime of approximately $1/f$ noise, in which further increases in the averaging time do not result in any further decrease in the type A uncertainty. The lowest type A uncertainty achievable with the ammeter, based on this data set, is $\sim 5$~fA, or 100 ppm of $I_{\text{A}}$. The electrometer, on the other hand, continues to follow $\sigma_{\text{A}} (\tau) \propto 1/ \sqrt{\tau}$ out to the longest time-scale probed by this data set, $\tau \sim 40000$~s, where $\sigma _{\text{A}} \sim 1$~fA, or 20 ppm of $I_{\text{E}}$.

Some insight into the behaviour of the ammeter can be gained by plotting the Allan deviation of a time-series of data taken with the instrument left open-circuit (open diamonds in figure \ref{AdevFig} (d)). This exhibits a transition to $1/f$ noise at $\tau \sim 10$~s, which is due to the low frequency behaviour of its input bias current noise. A small additional contribution may be due to the ADC voltage measurement \footnote{Instability in the analogue-to-digital conversion of the preamp output voltage can be ameliorated by using the ammeter's auto zero function. However, each reading then takes at least twice as long}. Referring to section \ref{SystSec}, the superior stability of the electrometer at long averaging times is probably a consequence of it having a more stable offset current and voltage than the ammeter. We might also propose that the electrometer has a more stable gain-setting element (the feedback capacitor) than the ammeter. However, in section \ref{AgreeSec}, and referring to the inset to figure \ref{AgreeFig} (a), we see that the ammeter gain is stable at the level of $10^{-5}$ on a time-scales of a few hour, so the $10^{-4}$ limit to the type A uncertainty discussed in the previous paragraph is unlikely to be due to instability in the resistive gain element.


\begin{figure}
\includegraphics[width=8.5cm]{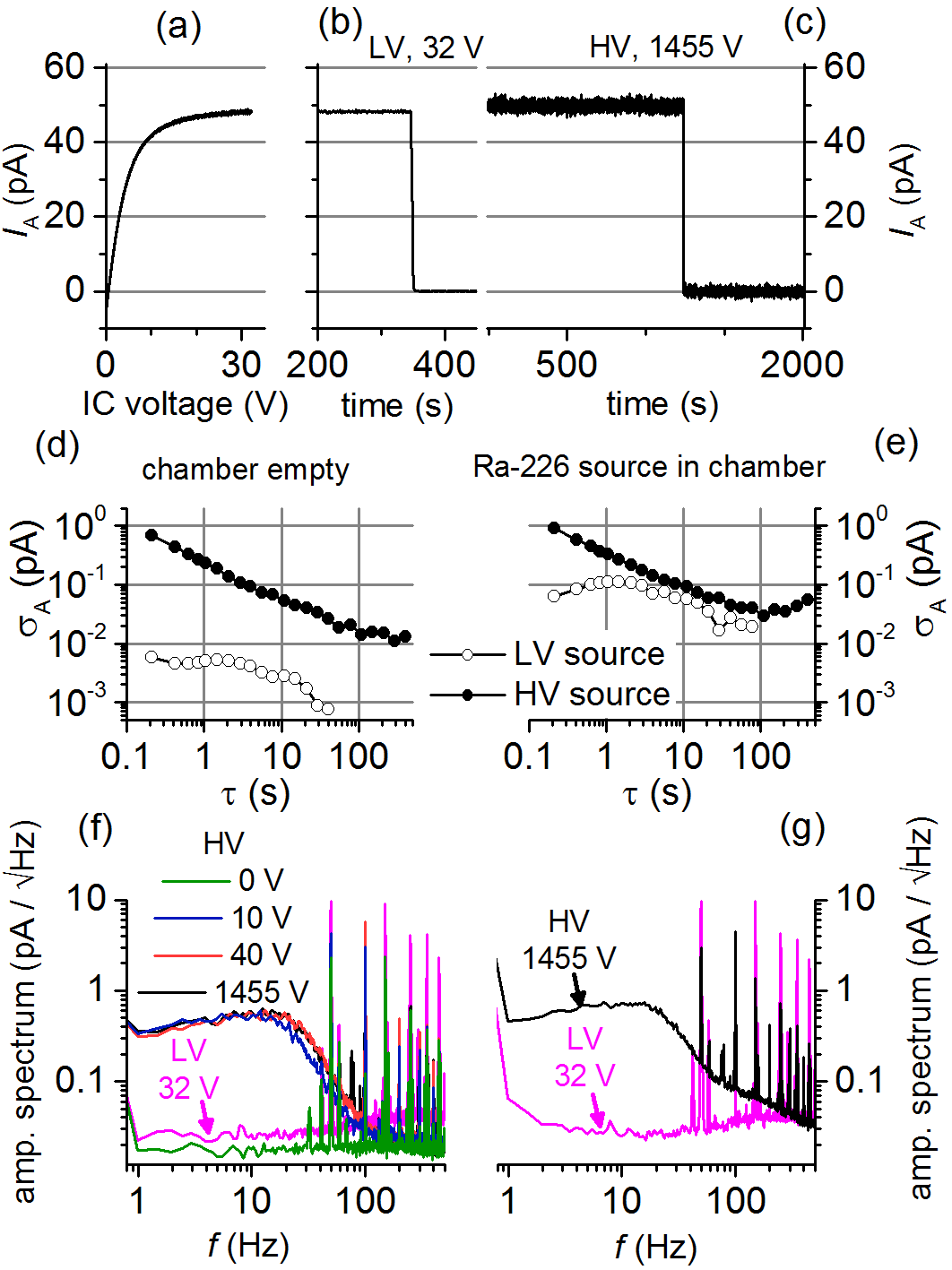}
\caption{\label{NoiseFig}\textsf{(a): Ammeter current as a function of ionisation chamber voltage, with the ionisation chamber energised with a low-noise laboratory DC supply. The Ra-226 source in the chamber is the same as in figure \ref{AdevFig}. (b,c): Ionisation chamber current with the chamber energised using (b): the low-voltage source and (c): the high-voltage source. In each data trace, the source is initially in the chamber, and is then removed. (d): Allan deviation of sections of data with the chamber empty from plots (b) and (c). Open symbols: LV source, filled symbols: HV source. (e): As (d), but with the Ra-226 source in the chamber. (f): Amplitude spectra of current noise from an empty chamber energised with the LV and HV sources. (g): as (f), but with the Ra-226 source in the chamber.}}
\end{figure}

The analysis presented in this section is not intended to be a definitive comparison of the two types of current measuring instrument, nor should the ammeter data be interpreted as definitively describing the particular make and model of instrument used in this study \footnote{The averaging time at which the Allan deviation changes from white-noise behaviour to frequency-dependent behaviour depends on the instrument range, the stability of the environmental conditions and the particular unit of instrument (of the same model number)}. Rather, it is intended to demonstrate a methodology for evaluating the type A uncertainty achieved following a given averaging time. For example, referring again to figure \ref{AdevFig} (d), if a statistical uncertainty of $50$~fA ($0.1 \%$ of the signal from the Ra-226 source) was desired, it is only necessary to integrate the current for $30$~s using either type of instrument. Knowledge of the stability of the current measuring instrument is also important when designing a protocol for measuring the chamber background current. One possible such protocol would be to measure the background current once a day, and subtract the same background from all calibrations performed that day. In this case, the Allan deviation of the readout current for $\tau = 1$~day would yield the minimum meaningful statistical uncertainty achievable in any calibration. Since instruments generally suffer from $1/f$ or random walk behavior at long time-scales, a more robust procedure would be to measure a new background signal every time the chamber is empty, i.e. in between calibrations of different sources.

\section{\label{NoiseSec} Investigation of excess noise}

A remarkable feature of the data in figure \ref{AdevFig} (d) is the roughly factor of $100$ increase in the short-averaging-time noise when the ammeter is connected to the energised ionisation chamber. This excess noise is indicated by a vertical double arrow. The excess noise is not due to the cable connecting the ammeter to the ionisation chamber. Separate measurements showed that the cable on its own, or indeed the cable connected to the chamber, but with the high voltage (HV) source disconnected from the chamber, increased the noise by a negligible amount compared to the situation with the ammeter input left open circuit. The statistical nature of current generation in the ionisation chamber can be expected to add a shot-noise contribution, but we do not believe this is a significant contributor to the total noise because there was only a small decrease in the total noise (less than a factor of $2$) when the source was removed from the chamber. 

To investigate the nature of the excess noise, we replaced the HV source with a low-noise laboratory voltage source (Yokogawa GS200), which we will refer to as the low-voltage (LV) source. This source was limited to a maximum of $32$~V, but as shown in figure \ref{NoiseFig} (a), the chamber current almost reached saturation at this voltage using the same Ra-226 source employed in the measurements reported in section \ref{TypeASec}. In figures \ref{NoiseFig} (b) and (c) we show data measured using the ammeter, in which the source was initially in the chamber, and was then withdrawn from the chamber. The data of figures \ref{NoiseFig} (b) and (c) were obtained using the LV and HV voltage sources, set to $32$~V and $1455$~V, respectively. The lower current noise when using the LV source is immediately apparent. Allan deviation plots of sections of the data from figures \ref{NoiseFig} (b) and (c), shown in panels (d) and (e) show, however, that the reduction in noise using the LV source is rather more complicated than might appear from the time-domain data plots. With the chamber empty, the reduction in noise using the LV source is indeed dramatic, at least a factor of 20 for averaging times from $0.2$~s to $100$~s. A single $0.2$~s data point using the LV source has a type A uncertainty of less than $10$~fA, while to achive the same type A uncertainty using the HV source requires averaging for at least $100$~s. With the Ra-226 source loaded into the chamber (figure \ref{NoiseFig} (e)), the LV source yields lower noise for averaging times up to a few seconds. For longer averaging times, the Allan deviation plots using the two voltage sources converge, and the LV source yields roughly a factor 2 lower noise than the HV source.

Next, we measured the amplitude spectra of the current noise using both the LV and HV sources, with the chamber empty and containing the Ra-226 check source. For these measurements, the Keithley 6430 ammeter was replaced with an ammeter setup consisting of a Femto DDPCA-300 transimpedance amplifier with gain set to $10^8$~V/A followed by a Keysight 34461A integrating voltmeter sampling $1000$ times a second. The bandwidth (3 dB point) of the transimpedance amplifier is $150$~Hz. Time-domain data traces were transformed in software to yield the amplitude spectra scaled in units of pA/$\sqrt{Hz}$ (figures \ref{NoiseFig} (f) and (g)). The spectra have peaks at multiples of the $50$~Hz power line frequency with both voltage sources, but the striking difference between the sources is at frequencies below about $50$~Hz, where the HV source generates a broad background with an amplitude more than ten times that of the LV source. The background due to the HV source persists even if its variable voltage is turned down as low as $10$~V, although it disappears if the voltage is set to zero. This data convincingly shows that the HV source is the origin of a large part of the the excess noise first seen in figure \ref{AdevFig} (d). We did not attempt to investigate the origin of the noise further, for example by directly measuring the voltage noise spectral density of the two voltage sources. It is nevertheless clear that elimination of excess noise due to the HV power supply, by filtering or improved design, would result in reductions in the amount of time required to achieve a given resolution in a measurement of ionisation chamber current, and more dramatic reductions in the time required to measure the background current. 

\section{\label{AgreeSec} Absolute agreement between two readout systems}

\subsection{Calibration of electrometer using reference current source}

\begin{figure}
\includegraphics[width=8.5cm]{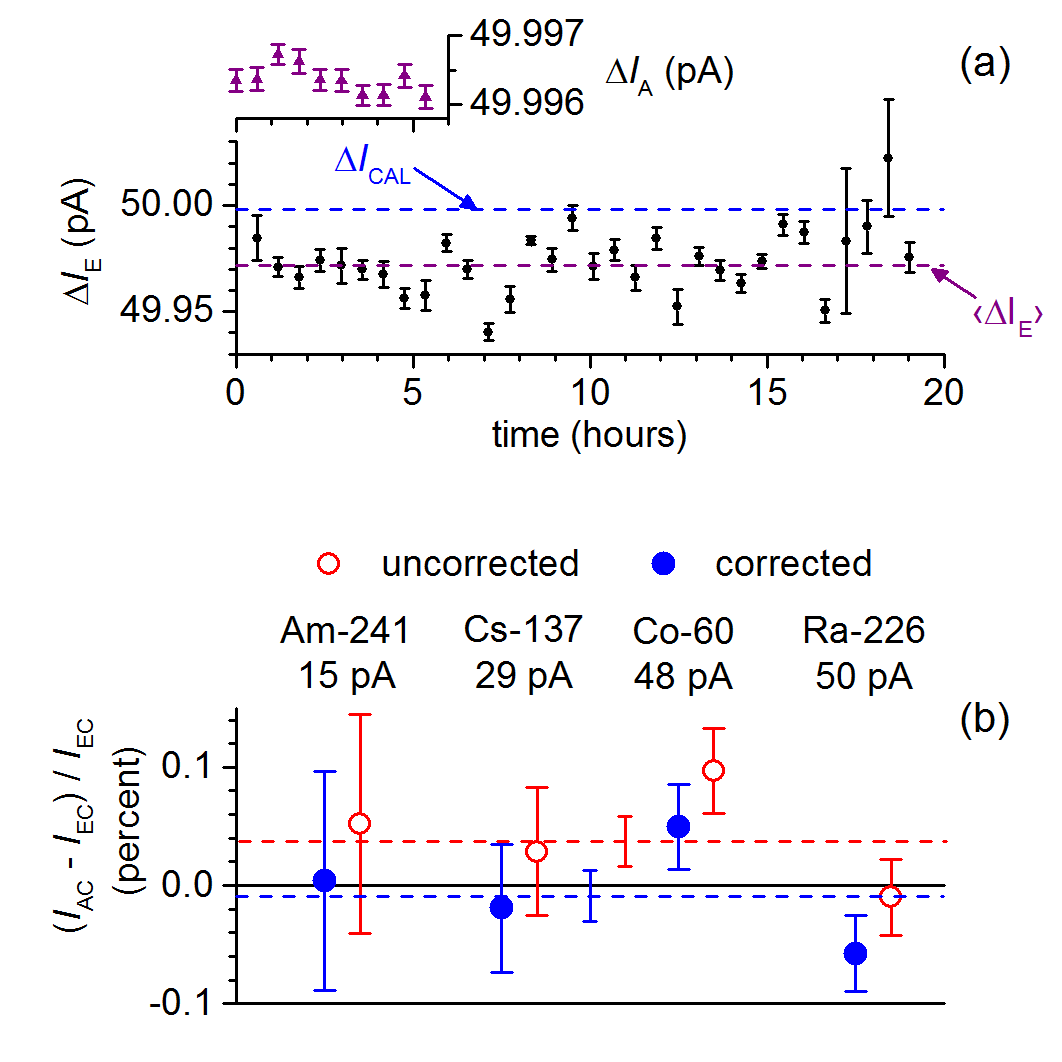}
\caption{\label{AgreeFig}\textsf{(a): The main plot shows the current indicated by the capacitor ramp electrometer when supplied with a known current of $49.995$~pA from a calibrated source. Each data point is averaged from $\sim 40$~minutes of voltage ramps, and the error bars indicate the standard error on the mean of the current calculated from the individual ramps. Horizontal dashed lines show the calibrated current (upper line) and the mean of the indicated currents (lower line). The upper left inset shows the current indicated by an ammeter when supplied with the same calibrated current. The inset shares the same time axis, and averaging time per point is the same as for the main plot, but note the different y-axis scales of the inset and main plot. (b): Agreement between the current measured by the electrometer and ammeter, when connected to an ionisation chamber in a series of measurements of four different radionuclides. Red open points: electrometer current as indicated. The red dashed line with error bar shows the weighted mean. Blue filled points: electrometer current corrected for the calibration factor determined from plot (a). The blue dashed line with error bar shows the weighted mean. Each measurement is corrected for background, and error bars indicate the type A uncertainty. The nuclide and approximate current are indicated above each pair of data points.}}
\end{figure}

We now return to the comparison between the ammeter and the electrometer. In this section, we investigate how well the two systems agree in background-corrected measurements of a range of radioactive sources. As already noted in section \ref{TracSec}, the gain factor of the ammeter was regularly calibrated using a reference current source consisting of a $1$~G$\Omega$ standard resistor and a calibrated DVM. Here, we also calibrated the gain factor of the electrometer using the same reference current source. For all the calibrations, the reference current was periodically switched between a nominal zero setting, and $50$~pA, yielding a difference current $\Delta I_{\text{cal}}=49.995$~pA. the difference currents $\Delta I_{\text{A}}$ and $\Delta I_{\text{E}}$ were extracted from the instrument readings. Figure \ref{AgreeFig} (a) shows values of $\Delta I_{\text{A}}$ (top-left inset) and $\Delta I_{\text{E}}$ (main plot) extracted from calibrations of the ammeter and electrometer respectively, over times of several hours. The most striking difference between the two instruments is that the values of $\Delta I_{\text{E}}$ exhibit much more statistical scatter than those for $\Delta I_{\text{A}}$ (note the different y-axis scales for the main panel of figure \ref{AgreeFig} (a) and the inset). This could be a consequence of the specialised design of the electrometer amplifier module: as discussed in section \ref{SystSec}, the input voltage noise of the amplifier module will cause excess noise when it is connected to the $1$~G$\Omega$ reference current source, and the electrometer may have a larger input voltage noise than the ammeter. However since we did not directly measure the voltage noise for either the electrometer or the ammeter, this remains a conjecture.

After averaging the statistical fluctuations in the calibration data of figure \ref{AgreeFig} (a), we find that the mean current difference indicated by the electrometer, $\langle \Delta I_{\text{E}} \rangle$, is offset from $\Delta I_{\text{cal}}$ by a statistically significant amount: $(\Delta I_{\text{cal}} - \langle \Delta I_{\text{E}} \rangle) / \Delta I_{\text{cal}} = (460 \pm 46) \times 10^{-6}$. This error, $460$ ppm, is much larger than the uncertainty in the capacitance, voltage and time components used to calculate $I_{\text{E}}$, although still much smaller than the uncertainties in the radionuclide-specific ionisation chamber calibration factors. We now consider the possible causes of this error.

The most likely cause of the error is non-linearity of the voltage ramp. In a previous study on another type of capacitor-ramp electrometer, non-linearity of the $V(t)$ ramp was at the level of a few parts in $10^{4}$ for currents in the range of $10$~pA to $100$~pA\cite{giblin2009si}. The non-linearity was assumed to arise due to dielectric storage, or other non-ideal properties of $C_{\text{stray}}$. However, it could not be satisfactorally modeled, and the measured non-linearity was used to assign empirical type B components to the uncertainty budget for the electrometer\cite{giblin2009si}. Measurements of $V(t)$ were also made on the electrometer under investigation in this study, and they also showed non-linearity at the level of a few parts in $10^{4}$. More extensive characterisation of the voltage ramp over a range of currents are needed to clarity this error mechanism.

As already noted in section \ref{TracSec}, a possible source of error in capacitor-ramp electrometers is frequency-dependence in the feedback capacitor. We measured the frequency dependence of the capacitor (a sealed-gas unit of $\sim 500$~pF) over the range $50-20000$~Hz using a commercial precision capacitance bridge (AH2700A), and found that it only changed by a few ppm. However, we did not measure the capacitance at the millihertz frequencies at which the electrometer operates. In a previous study, it was found that capacitors with a large frequency dependence in the millhertz range also showed an anomalously large dependence in the audio range \cite{giblin2010frequency}. However, this study was based on a small sample of capacitors, and we cannot conclusively rule out capacitor frequency dependence as the cause of the $460$~ppm gain error in the ionisation chamber electrometer.

Finally, the error may be simply an artifact due to the input resistance of the electrometer in conjunction with the $1$~G$\Omega$ output resistance of the reference current source. The sign of the error (the indicated current is less than the actual current) is consistent with this mechanism. The measured $460$~ppm error would imply an input resistance of $460$~k$\Omega$, which is quite high but not implausible. Future calibrations, using reference current sources with different output resistances, will clarify this matter. In the following comparison between the ammeter and electrometer, we simply treat the calibration of the electrometer as yielding a correction factor, in the same manner in which we calibrated the ammeter.

\subsection{Background-corrected measurements using both readout systems}

As a direct comparison, the electrometer and ammeter were both used to measure background-corrected ionisation chamber currents from four different radionuclides. Each measurement consisted of a raw data set similar to that shown in figure \ref{NoiseFig} (c), from which the background corrected currents $I_{\text{AC}}$ and $I_{\text{EC}}$ were obtained. To ensure that random geometrical factors due to source placement inside the chamber did not affect the comparison, the source was only put into the chamber once for each comparison. So, for example the ammeter would be used to measure first the empty chamber, then the source. Next the electrometer would be used to measure the source followed by the empty chamber. The socket at which the instruments were connected and disconnected from the chamber was mechanically isolated from the chamber via a cable to avoid disturbing the position of the source when the instruments were swapped. As detailed, $I_{\text{AC}}$ already incorporates a correction factor from the ammeter calibration. $I_{\text{EC}}$ was optionally corrected, based on the calibration detailed in the previous sub-section. Figure \ref{AgreeFig} (b) shows the normalised difference between the two background corrected currents both with and without the calibration correction applied to the electrometer current. After applying the correction, the weighted mean of the $4$ points yields the average $\langle \frac{I_{\text{AC}}-I_{\text{EC}}}{I_{\text{EC}}} \rangle = (-0.009 \pm 0.021) \%$, as indicated by the blue horizontal dashed line and error bar; the two systems agree within the random uncertainties. Without applying the correction factor, the weighted mean of the normalised differences is $(0.037 \pm 0.021) \%$, a statistically significant disagreement. The measurement of the 4 radionuclides could be considered as an indirect comparison of the two current measuring instruments, although with a higher uncertainty than the direct calibrations discussed in the previous section. Our ability to compare the two measurement systems with radionuclide measurements is hampered by the excess noise, probably due to the HV supply as discussed in section \ref{NoiseSec}, but we conclude that once they are both calibrated using a reference current, they agree to within $\sim 0.02 \%$. They could be considered as equivalent candidates for an ionisation chamber readout system, provided a reference current source was available to calibrate them.

\section{\label{ConcSec} Conclusions}

We compared examples of a feedback ammeter and an integrating electrometer, and we can conclude that the feedback ammeter, calibrated using a reference current source, can be considered as a viable alternative to the integrating electrometer traditionally used for ionisation chamber readout. Measuring ionisation chamber currents of a few tens of picoamps at an uncertainty level of $0.1 \%$, which is sufficient for most radionuclide calibrations, the two current readout systems can be considered equivalent. At an uncertainty level of $0.01 \%$, the two systems can also be considered equivalent with respect to type A uncertainty, reaching a relative type A uncertainty of $0.01 \%$ for a current of $50$~pA after $1000$ seconds of averaging. However, when calibrated using a reference current source, the electrometer was found to be in error by $0.046 \%$. This highlights the importance of calibrating electrometers directly using reference current sources, as non-idealities in these systems can introduce errors orders of magnitude larger than the ppm-level uncertainties in the individual calibrations of capacitance, voltage and time. Reference current sources can be realised at uncertainty levels of around 1 ppm using calibrated standard resistors, voltmeters, and now the ULCA.

Independent of the readout system, the type A uncertainty was increased by a significant amount above the measuring instrument noise floor by a large amount of background noise originating in the high-voltage source. This shows that careful engineering of a low-noise high-voltage source would be a fruitful project, enabling type A uncertainties less than $0.01 \%$ to be achieved in just a few seconds of measurement time. We have also presented the Allan deviation as a useful statistical tool for evaluating the stability of current measuring instruments as a function of measuring time. This helps the design of calibration protocols which make most efficient use of the available time to reach a desired uncertainty level. 

\begin{acknowledgments}
This research was supported by the UK department for Business, Energy and Industrial Strategy and the EMPIR Joint Research Project 'e-SI-Amp' (15SIB08). The European Metrology Programme for Innovation and Research (EMPIR) is co-financed by the Participating States and from the European Union's Horizon 2020 research and innovation programme. The authors would like to thank Kelley Ferreira for assistance with handling radionuclide samples.
\end{acknowledgments}

\bibliography{SPGrefs_IonChamberPaper}

\end{document}